\documentclass[a4paper,12pt]{article}
\usepackage{graphicx,amssymb,bm,latexsym,color,epsf,cite}
 \usepackage{amsmath}
 \usepackage{appendix}
\makeatletter
\@addtoreset{equation}{section}

\makeatother
\newcommand{\bea}{\begin{eqnarray}}
\newcommand{\eea}{\end{eqnarray}}
\newcommand{\vs}[1]{\vspace{#1 mm}}
\newcommand{\hs}[1]{\hspace{#1 mm}}
\renewcommand{\a}{\alpha}
\renewcommand{\b}{\beta}

\renewcommand{\d}{\delta}
\newcommand{\e}{\epsilon}
\newcommand{\s}{\sigma}

\newcommand{\la}{\lambda}
\newcommand{\pa}{\partial}
\newcommand{\nn}{\nonumber\\}
\newcommand{\p}[1]{(\ref{#1})}

\newcommand{\tlh}{\tilde h}
\newcommand{\Propag}[1]{{\rm F.T.}\langle0|\,{\rm T}\, #1 \,|0\rangle}
\newcommand{\VEV}[1]{\langle 0| #1 |0\rangle}
\newcommand{\B}{{\rm B}}
\newcommand{\GF}{{\rm GF}}
\newcommand{\FP}{{\rm FP}}
\newcommand{\T}{{\rm T}}

\begin{document}

\begin{flushright}
YITP-21-87 \\
\end{flushright}
\medskip
\renewcommand{\thefootnote}{\fnsymbol{footnote}}

\begin{center}
{\Large\bf
BRST Quantization of General Relativity in Unimodular Gauge and Unimodular Gravity
}
\vs{5}

{\large
Taichiro Kugo,$^{1,}$\footnote{e-mail address: kugo@yukawa.kyoto-u.ac.jp}
Ryuichi Nakayama,$^{2,}$\footnote{e-mail address: nakayama@particle.sci.hokudai.ac.jp}
and
Nobuyoshi Ohta\,$^{3,4,}$\footnote{e-mail address: ohtan@ncu.edu.tw}
} \\
\vs{5}

$^1$
{\em Yukawa Institute for Theoretical Physics, Kyoto University, Kyoto 606-8502, Japan}
\vs{3}

$^2$
{\em Division of Physics, Graduate School of Science,
Hokkaido University, Sapporo 060-0810, Japan}
\vs{3}

$^3$
{\em Department of Physics, National Central University, Zhongli, Taoyuan 320317, Taiwan}

and

$^4$
{\em Research Institute for Science and Technology,
Kindai University, Higashi-Osaka, Osaka 577-8502, Japan
}

\vs{5}
{\bf Abstract}
\end{center}

Unimodular gravity (UG) is an important theory which may explain the smallness of the cosmological constant.
To get insight into the covariant quantization of UG, we discuss the BRST quantization of
General Relativity (GR) with a cosmological constant in the unimodular gauge.
We develop a novel way to gauge fix the transverse diffeomorphism (TDiff) and then further
to fulfill the unimodular gauge. This process requires the introduction of an additional pair of
BRST doublets which decouple from the physical sector together with the other three pairs of BRST doublets
for the TDiff. We show that the physical spectrum is the same as GR in the usual covariant gauge fixing.
We then study a theory derived by making ``Fourier transform'' of GR in the unimodular gauge with respect to
the cosmological constant as a candidate of ``quantum UG.''
We clarify the difference from GR and point out problems in this theory.

\renewcommand{\thefootnote}{\arabic{footnote}}
\setcounter{footnote}{0}

\section{Introduction}

General Relativity (GR) well describes gravitational low-energy phenomena.
The action is given by
\bea
S_{\rm GR}= Z_N \int d^4 x \sqrt{-g} \left(R -2 \Lambda \right),
\label{GRaction}
\eea
where $Z_N =\frac{1}{16\pi G_N}$ with the Newton constant $G_N$.
Here we have also included the cosmological constant, and we refer to this theory simply as GR.
Making variation of this theory, we get the Einstein equation with the cosmological constant:
\bea
R_{\mu\nu}-\frac12 (R-2\Lambda) g_{\mu\nu}=0.
\label{ein}
\eea

Classical GR has several equivalent formulations.
An important example is the Unimodular Gravity (UG)~\cite{BD1,BD2,Unruh,HT,EVMU,NV},
which is defined as a theory of gravity with the constraint
\bea
\sqrt{-g}=\omega,
\label{uni}
\eea
where $\omega$ is a fixed volume form. This constraint can be imposed by using a Lagrange multiplier field $\lambda$:
\bea
S_{\rm UG}= Z_N \int d^4 x \left[ \sqrt{-g} (R -2 \Lambda) + \la(\sqrt{-g}-\omega)\right].
\label{UGaction}
\eea

This theory is considered to give a solution to the cosmological constant problem.
Making the variation of the action with respect to the metric, we obtain
\bea
R_{\mu\nu}-\frac{1}{2}(R-2\Lambda+\la)g_{\mu\nu}=0,
\label{fe}
\eea
together with the unimodular condition~\p{uni}.
Taking the trace of this equation, we get
\bea
\la=2\Lambda-\frac{R}{2}.
\label{lambda}
\eea
Plugging this into the field equation~\p{fe}, we find
\bea
R_{\mu\nu}-\frac{1}{4}R g_{\mu\nu}=0,
\label{ugfe}
\eea
which is the traceless part of the Einstein equation.
Thus the ``cosmological constant'' drops out of our field equation.
In this theory, we cannot get the trace part of the usual Einstein equation, but this can be recovered
using the Bianchi identity.
By taking the covariant derivative of \p{ugfe}, we get
\bea
\pa_\mu R=0,
\eea
which, upon integration yields
\bea
R=4\Lambda_0 .
\eea
Then eq.~\p{ugfe} can be rewritten as
\bea
R_{\mu\nu}-\frac{1}{2}Rg_{\mu\nu}+\Lambda_0\, g_{\mu\nu}=0,
\eea
recovering the Einstein equation with the cosmological constant $\Lambda_0$.
Thus the cosmological constant appears as an integration constant.
Note that $\Lambda_0$ has nothing to do with the constant term in the action.
In the presence of matter, this could be repeated together with the conservation of energy-momentum tensor.

The constant term $\Lambda$ in \p{UGaction} can be absorbed into the definition of the Lagrange multiplier field $\la$:
\bea
S_{\rm UG}= Z_N \int d^4 x \left[ \sqrt{-g} R + \la(\sqrt{-g}-\omega)\right].
\label{UGaction2}
\eea
Using the unimodular condition, our action can also be written as
\bea
S_{\rm UG}= Z_N \int d^4 x \left[ \omega R + \la(\sqrt{-g}-\omega)\right].
\label{UGaction3}
\eea
This is the standard formulation of UG, and this shows that the classical gravity
is equivalent to UG modulo the above-mentioned global property.

At the classical level, GR and UG are almost equivalent in the sense that \p{uni} can be seen
as a gauge fixing condition in GR.
However when this condition is present, the identity for the global quantity,
\bea
\int d^4 x \sqrt{-g} =\int d^4 x\, \omega,
\label{vol}
\eea
follows, which is a gauge (GC transformation) invariant relation and cannot be regarded merely as a gauge condition.
Thus classical UG may be considered to be equivalent to the classical GR with the fixed spacetime volume.
Conversely GR may be considered to be obtained by summing over the spacetime volume.

The fact that the ``cosmological constant'' in the action does not have any physical meaning is
an important advantage of UG. If we consider the history of the universe, our universe underwent
phase transitions before it settles down to the spontaneous broken phase of various symmetries including
the electroweak symmetry, and this must have produced huge vacuum energy. In addition, the quantum fluctuation
of the gravity and matter also produces huge vacuum energy. Both must be canceled out by fine tuning
in GR, but it is not necessary in UG.

The price we have to pay is that the action~\p{UGaction2} or \p{UGaction3} is no longer invariant under
the full diffeomorphism, but only under the transverse (or, volume-preserving) diffeomorphism (TDiff):
\bea
\d g_{\mu\nu} = \nabla_\mu \e_\nu^T +\nabla_\nu \e_\mu^T, \qquad
\nabla^\mu \e_\mu^T=0.
\eea
This gives a source of confusion as to the question
if this theory is equivalent to GR or not at the quantum level.
At first sight, since this theory could be regarded as just a partially gauge fixed theory of GR,
one would expect that it is equivalent to GR, but there has been lots of debate on
the equivalence~\cite{Smolin,FG,Eichhorn2013,Saltas,PS,Alvarez0,Alvarez1,BOT,Percacci2017,DOP,GM,HS,BP,DMPP,B,N}.
In fact upon further consideration, this may not be so trivial. Since the theory has invariance
only under TDiff, we expect that this invariance removes
degrees of freedom one less than the full diffeomorphism. This reduction of degrees of freedom
could be done by the Faddeev-Popov (FP) ghosts. In GR,
we have the full diffeomorphism with 4 coordinate parameters, and this introduces 4 sets of
FP ghosts and antighosts, leaving $10-8=2$ degrees of freedom, properly.
In UG, however, the gauge fixing can be done only for the transverse modes, with 3 sets of
ghosts. The additional condition is provided by the unimodularity condition~\p{uni}, but
this condition does not seem to require Faddeev-Popov ghosts.
This leaves the question how the remaining degrees of freedom are removed.

On the other hand, we can use the reparametrization invariance in GR to partially gauge fix to satisfy
the unimodular gauge condition~\p{uni}. It is then an interesting problem how the above problem on degrees
of freedom is resolved in this gauge, since the remaining reparametrization invariance is only
TDiff. Once understood properly, this is expected to cast light on how to quantize UG.

In this respect, Baulieu \cite{B} has recently proposed an interesting way of
gauge fixing which seems to realize this unimodular gauge in a manifestly covariant manner.
He introduces a new set of ``BRST quartet'' fields consisting of an additional scalar field together with
ghosts and a scalar. The reasoning is, however, not very clear and it is difficult to understand
the nature and the role of the added scalar field as well as the origin of the additional  BRST symmetry.

Here we propose a novel and general way of decomposing a $d$-vector condition into a scalar plus $d-1$
(i.e., transverse-vector) conditions in a manifestly covariant and local manner,
which naturally leads to the introduction of a scalar field and the emergence of a new gauge symmetry.
This reproduces essentially the same unimodular gauge-fixed Lagrangian as Baulieu's one.
Naively one might expect that the newly introduced four fields would be a ``BRST quartet'' and decouple
by themselves. We find that this is not the case, and the BRST multiplets rearrange themselves in a very subtle way.
We show that after all they completely decouple and this covariant theory gives the correct number of degrees
of freedom in GR.

Our key observation is that when we gauge fix TDiff, we have to deliberately impose
the gauge condition such that we impose it only on the transverse modes.
In doing this, we introduce an additional longitudinal mode and then there appears additional invariance,
which removes the remaining degrees of freedom.
Since this is the quantization of GR, the cosmological constant is present in the field equation even though
we fix the theory to unimodular gauge with $\sqrt{-g}=\omega$.
We expect that our new way of describing GR in the unimodular gauge is useful for understanding
the formulation of quantum theory of UG. Our discussion is valid in the presence of matter fields.

We then discuss what implications this formulation of GR may have to the problem of quantization of UG.
The unimodularity condition~\p{uni} is imposed as a constraint in UG whereas it is just a gauge choice in GR.
Similarly the spacetime volume is fixed in UG whereas the cosmological constant is fixed in GR.
Considering the dual role of these~\cite{Hawking}, it would be natural to consider that the quantum theory of UG
may be obtained by the Fourier transform of GR in the unimodular gauge with respect to the cosmological constant.
Unfortunately we find that such a formulation has problem with unitarity, but we hope that
our discussions clarify the problems in the covariant quantization of UG.

\section{BRST quantization of GR in the unimodular gauge}

\subsection{Covariant unimodular gauge fixing in GR}

Now let us consider the unimodular gauge fixing in GR.
Using the reparametrization invariance, we can partially gauge fix the theory to the unimodular gauge
with the condition~\p{uni}.

In order to quantize the theory, we have to gauge fix the theory.
In the usual GR, we introduce the usual gauge fixing condition called
de Donder gauge:
\bea
f^\mu \equiv \pa_\rho \tilde g^{\rho\mu}=0.
\qquad (\tilde g^{\mu\nu}\equiv \sqrt{-g} g^{\mu\nu}).
\label{gf1}
\eea
This contains $d=4$ conditions and completely fixes all the diffeomorphism transformations.
In our theory where we will take the unimodular gauge condition~\p{uni} as a partial gauge fixing, however,
Eq.~\p{gf1} imposes too many conditions, and we have to remove one condition from this.
If we gave up manifest covariance, we could take only $d-1=3$
conditions for the spacial components $f^i=0$ $(i=1,2,3)$ of the de Donder
gauge ~\p{gf1}. However, we want to keep manifest covariance here.

Now comes our first crucial step.
Instead of imposing the $d$-vector condition~\p{gf1}, we introduce an additional field $S$ and impose the condition
\bea
\tilde g_{\mu\nu} f^\nu =\pa_\mu S.
\label{gf2}
\eea
The meaning of this gauge fixing is as follows:
The right-hand side (rhs) of this equation is a derivative of the scalar field $S$, and so
the condition~\p{gf2} requires the transverse component of the left-hand side (lhs) should vanish.
The longitudinal component simply defines the scalar field $S$, which is left arbitrary.
It is important that we consider full diffeomorphism transformation here, but this
condition~\p{gf2} imposes the gauge condition only on the transverse modes in $\tilde g_{\mu\nu} f^\nu$.

Our second crucial observation is the following.
Our unimodularity condition means that the remaining longitudinal mode should be removed from the spectrum
of the theory. We then notice that the new field $S$ does not appear at all
in the original action of our GR theory, and hence the action is trivially
invariant under an arbitrary shift of the scalar field $S$:
\bea
\d S(x) = d(x).
\label{inv}
\eea
Note that this is a (hidden) gauge-invariance. (Such a gauge-invariance under
the arbitrary change of the fields not appearing in the action, is trivial but
was recognized useful by Izawa~\cite{Izawa} in the context of BRST gauge-fixing.)
We can lift this new gauge invariance as well as the original GC transformation invariance to those
under the BRST transformation defined by
\bea
\d_\B g_{\mu\nu} &=& g_{\mu\rho} \pa_\nu c^\rho + g_{\nu\rho}\pa_\mu c^\rho+ c^\rho \pa_\rho g_{\mu\nu}, \nn
\d_\B c^\mu &=& c^\rho \pa_\rho c^\mu, \nn
\d_\B \bar c_\mu &=& i b_\mu, \qquad
\d_\B b_\mu =0, \\
\d_\B S &=& d, \qquad
\d_\B d = 0, \nn
\d_\B \bar d &=& i b, \qquad
\d_\B b =0,\nonumber
\eea
where $c^\mu$ and $\bar c_\mu$ are the anticommuting ghosts corresponding to the diffeomorphism transformations,
$b_\mu$ and $b$ are the Nakanishi-Lautrup (NL) fields~\cite{Nakanishi:1966zz},
and $d$ and $\bar d$ are anticommuting ghosts corresponding to the invariance~\p{inv}.

Given the invariance and BRST transformation, we have a systematic method to give the gauge fixing and
ghost terms~\cite{KU,Ohta2020}. Thus our gauge-fixing and ghost terms are given by
\begin{eqnarray}
{\cal L}_{\GF+\FP}&=& -i \delta_\B \Bigl[
\bar c_\mu \Bigl( \partial_\nu\tilde g^{\mu\nu}
- \tilde g^{\mu\nu}\partial_\nu S +\frac{\a}{2} \eta^{\mu\nu}b_\nu
 \Bigr)
+\bar d\left( \sqrt{-g} -\omega +\frac{\b}{2}b \right)
\Bigr] \nonumber \\
&=&
b_\mu \Bigl( \partial_\nu\tilde g^{\mu\nu} - \tilde g^{\mu\nu}\partial_\nu S \Bigr)
+ \frac{\a}{2} \eta^{\mu\nu}b_\mu b_\nu
+b\left( \sqrt{-g} -\omega  \right)+\frac{\b}{2}b^2  \nonumber \\
&&+i\bar c_\mu \Bigl( \partial_\nu\delta_\B(\tilde g^{\mu\nu}) - \delta_\B(\tilde g^{\mu\nu})\partial_\nu S
 - \tilde g^{\mu\nu}\partial_\nu d \Bigr)
+i \bar d \left( \sqrt{-g}\nabla_\mu c^\mu \right),
\label{GR:gf}
\end{eqnarray}
where BRST transform of $\tilde g^{\mu\nu} = \sqrt{-g} g^{\mu\nu}$ is
\begin{equation}
\delta_\B(\tilde g^{\mu\nu})=
-\partial_\rho c^\mu\cdot\tilde g^{\rho\nu}-\partial_\rho c^\nu\cdot\tilde g^{\mu\rho}
+\partial_\rho(c^\rho\tilde g^{\mu\nu}),
\end{equation}
and $\a$ and $\b$ are gauge parameters, and $\omega$ is an arbitrary gauge function fixing the `unimodularity'.
We see that the first term in \p{GR:gf} only fixes the transverse coordinate transformations,
and $d$ and $\bar d$ restricts the reparametrization ghost $c^\mu, \bar c_\mu$ to transverse components.
The additional set of fields $(S, d, \bar d, b)$ apparently looks to form a BRST quartet
but actually they do not. We are now going to show that they rearrange with other ghosts
into sets of BRST quartets and all of them completely decouple.

\subsection{Propagators}

Let us check in more detail if we get nonsingular fully gauge fixed action with the correct degrees of freedom
on the flat background with $\omega=1$. For the GR theory with $\Lambda=0$,
setting the gauge parameters $\a=\b=0$ for simplicity, our total action is given as
\begin{eqnarray}
{\cal L}_{\rm GR}
&=& \sqrt{-g}R +
b_\mu \Bigl( \partial_\nu\tilde g^{\mu\nu} - \tilde g^{\mu\nu}\partial_\nu S \Bigr)
+b\left( \sqrt{-g} - 1 \right)   \nonumber \\
&&+i\bar c_\mu \Bigl( \partial_\nu\delta_\B(\tilde g^{\mu\nu}) - \delta_\B(\tilde g^{\mu\nu})\partial_\nu S
 - \tilde g^{\mu\nu}\partial_\nu d \Bigr)
+i \bar d \left( \sqrt{-g}\nabla_\mu c^\mu \right).
\end{eqnarray}
The fluctuation is defined by
\bea
\tilde g^{\mu\nu} = \eta^{\mu\nu} +\tlh^{\mu\nu},
\eea
with $\eta_{\mu\nu}= \mbox{diag}(-1, +1,+1,+1)$ and then to the linear order we have
\bea
g_{\mu\nu} = \eta_{\mu\nu} -\tlh_{\mu\nu}+\frac12 \eta_{\mu\nu} \tlh + \cdots,
\qquad
\sqrt{-g} = \sqrt{-\tilde g}= 1+\frac12 \tlh +\cdots,
\eea
where $\tlh\equiv \tlh_\mu^\mu$ is the trace of the fluctuation.

The quadratic terms of ${\cal L}_{\rm GR}$ are given by
\begin{eqnarray}
{\cal L}_{\rm GR}\Bigr|_{\rm quadr} &=&
{\cal L}_{\rm boson} + {\cal L}_{\rm ghost}, \nn
{\cal L}_{\rm boson}
&=& \frac14 \tlh_{\mu\nu}\square \tlh^{\mu\nu}+\frac12 (\pa_\nu \tlh^{\mu\nu})^2
-\frac18 \tlh\square \tlh
+b_\mu (\pa_\nu \tlh^{\mu\nu}-\pa^\mu S)  +\frac12 b\tlh, \nn
{\cal L}_{\rm ghost}&=&
i \bar c_\mu [\pa_\nu(-\pa^\nu c^\mu-\pa^\mu c^\nu+ \pa_\rho c^\rho \eta^{\mu\nu}
)-\pa^\mu d]
+ i\bar d\pa_\mu c^\mu \nn
&=& i \bar c_\mu [-\square c^\mu-\pa^\mu d] + i\bar d \pa_\mu c^\mu,
\end{eqnarray}
which are written in the matrix form 
\begin{eqnarray}
\frac12 \bigl(
\tlh_{\mu\nu}, S, b_\mu, b \bigr)
\Gamma^{(2)}_{\rm boson}
\begin{pmatrix}
\tlh_{\rho\sigma} \\
S \\
b_\rho\\
b
\end{pmatrix}
+
(\bar c_\mu, \bar d)
\Gamma^{(2)}_{\rm ghost}
\left( \begin{array}{c}
c^\nu \\d
\end{array}\right),
\end{eqnarray}
in terms of the 2-point vertex matrices
$\Gamma^{(2)}_{\rm boson}$ and
$\Gamma^{(2)}_{\rm ghost}$, the inverses of which give the propagators.
If the gauge is fixed properly by the present gauge, then the
inverses exist and the propagators are determined unambiguously.

Let us begin with the simpler ghost sector. The 2-point vertex
$\Gamma^{(2)}_{\rm ghost}$ in momentum space reads
\begin{eqnarray}
\Gamma^{(2)}_{\rm ghost} =
 -i \times \bordermatrix{
         &   c^\nu           & d     \cr
\bar c_\mu&  -p^2\delta^\mu_\nu  &  ip^\mu \cr
\bar d   &  -ip_\nu         &  0    \cr
},
\end{eqnarray}
the inverse of which surely exists and is given by
\begin{equation}
{\Gamma^{(2)}}^{-1}_{\rm ghost}=
i\frac1{-p^2}\times
\bordermatrix{
         &   \bar c_\nu         &  \bar d      \cr
c^\mu& \delta^\mu_\nu-p^\mu p_\nu /p^2  &  -ip^\mu \cr
 d   &  ip_\nu         &  -p^2   \cr
}.
\label{eq:GammaInvGhost}
\end{equation}
That is, we find the ghost propagators $i\times {\Gamma^{(2)}}^{-1}_{\rm ghost}$:
\begin{eqnarray}
\Propag{c^\mu\,\bar c_\nu} &=&
\frac{ \delta^\mu_\nu - \frac{p^\mu p_\nu}{p^2}}{p^2}, \\
\Propag{c^\mu\,\bar d} &=& \frac{ -ip^\mu}{p^2}, \qquad
\Propag{d\, \bar c_\nu} = \frac{ ip_\nu }{p^2}, \\
\Propag{ d \,\bar d} &=& -1\,,
\end{eqnarray}
where the F.T. means
\begin{equation}
\Propag{\phi_1\, \phi_2} \equiv
\int d^4x e^{-ip(x-y)} \VEV{{\rm T} \phi_1(x)\, \phi_2(y)}\,.
\end{equation}

Next is the boson sector:
We find the 2-point vertex $\Gamma^{(2)}_{\rm boson}$ in momentum space
\begin{eqnarray}
&&\hspace{-1em}\Gamma^{(2)}_{\rm boson} = \nonumber \\
&&\bordermatrix{
           & \tlh_{\rho\sigma}  & S & b_\rho& b  \cr
\tlh_{\mu\nu} &
\begin{matrix}
-p^2\,\Bigl[\frac12 I^{\mu\nu,\rho\sigma}-\frac1{12}d^{\mu\nu}d^{\rho\sigma}
\phantom{\frac34e^{\mu\nu}e^{\rho\sigma}\frac34e^{\mu\nu}e^{\rho\sigma}}
\\
-\frac14\left(d^{\mu\nu}e^{\rho\sigma}+e^{\mu\nu}d^{\rho\s}\right)
-\frac34e^{\mu\nu}e^{\rho\sigma}\Bigr]
\end{matrix}
       & 0 & -i\eta^{\rho(\mu} p^{\nu)} & \frac12\eta^{\mu\nu}
         \cr
S & 0  & 0 & ip^\rho              &  0 \cr
b_\mu& i\eta^{\mu(\rho}p^{\sigma)}  &-ip^\mu & 0 & 0 \cr
b & \frac12 \eta^{\rho\sigma}  & 0&0 & 0 \cr
},~~~
\end{eqnarray}
by using the projection operators
\def\II{{I\kern-2.5ptI}}
\begin{eqnarray}
&&d_{\mu\nu} = \eta_{\mu\nu}- \frac{p_\mu p_\nu}{p^2}, \qquad
e_{\mu\nu}= \frac{p_\mu p_\nu}{p^2},  \\
&&I_{\mu\nu,\rho\sigma}= \frac12\left(
d_{\mu\rho}d_{\nu\sigma}+d_{\mu\sigma}d_{\nu\rho}-\frac23 d_{\mu\nu}d_{\rho\sigma}\right).
\end{eqnarray}
Noting their projection properties
\begin{eqnarray}
&&p^\mu d_{\mu\nu}=0,\quad d_{\mu\nu}\eta^{\mu\nu}=3,\quad e_{\mu\nu}\eta^{\mu\nu}=1, \\
&&
d_{\mu\alpha}d^{\alpha\nu}=d_\mu{}^\nu,\quad e_{\mu\alpha}e^{\alpha\nu}=e_\mu{}^\nu,
\quad d_{\mu\alpha}e^{\alpha\nu}=0, \\
&&I_{\mu\nu,\alpha\beta}d^{\alpha\beta}= 0,\quad
I_{\mu\nu,\alpha\beta}e^{\alpha\beta}= 0,\quad
I_{\mu\nu,\alpha\beta}I^{\alpha\beta,\rho\sigma}= I_{\mu\nu}{}^{\rho\sigma},
\end{eqnarray}
we can straightforwardly compute the inverse of this matrix,
${\Gamma^{(2)}}^{-1}_{\rm boson}$,
with a little effort:
\begin{eqnarray}
&&\hspace{-1em}{\Gamma^{(2)}}^{-1}_{\rm boson}=
\frac1{-p^2}\times
\nonumber \\
&&\bordermatrix{
           & \tlh_{\rho\sigma}  & S & b_\rho& b  \cr
\tlh_{\mu\nu} &
\begin{matrix}
\Bigl[ 2 I^{\mu\nu,\rho\sigma}-\frac13 d^{\mu\nu}d^{\rho\sigma}
\phantom{\frac34e^{\mu\nu}e^{\rho\sigma}\frac34e^{\mu\nu}e^{\rho\sigma}}
\\
+\left(d^{\mu\nu}e^{\rho\sigma}+e^{\mu\nu}d^{\rho\sigma}\right)
-3 e^{\mu\nu}e^{\rho\sigma}\Bigr]
\end{matrix}
    & d^{\mu\nu} - 3e^{\mu\nu} & 2i p^{(\mu}d^{\nu)\rho}  & -p^2(d^{\mu\nu}-e^{\mu\nu})  \cr
S & d^{\rho\sigma} -3 e^{\rho\sigma} & -3  & -ip^\rho           &  p^2 \cr
b_\mu& -2ip^{(\rho}d^{\sigma)\mu}  &ip^\mu & 0 & 0 \cr
b & -p^2 ( d^{\rho\sigma}-e^{\rho\sigma})  & p^2 & 0 & 0 \cr
}. \nn
\label{eq:GammaInvBoson}
\end{eqnarray}
Thus, we explicitly confirmed that the inverse also exists
in the boson sector and hence that our gauge fixing is complete.
The propagators given by $i \times {\Gamma^{(2)}}^{-1}_{\rm boson}$
read e.g.,
\begin{eqnarray}
\Propag{\tlh^{\mu\nu}\,\tlh^{\rho\sigma}} &=&
i\, \frac{
2 I^{\mu\nu,\rho\sigma}-\frac13 d^{\mu\nu}d^{\rho\sigma}
+\left(d^{\mu\nu}e^{\rho\sigma}+e^{\mu\nu}d^{\rho\sigma}\right)
-3 e^{\mu\nu}e^{\rho\sigma} }{-p^2}, \nn
\Propag{\tlh^{\mu\nu}\,S} &=&
i\, \frac{d^{\mu\nu}-3e^{\mu\nu}}{-p^2},
\end{eqnarray}
etc. Note that the graviton propagator has not only a dipole but also a tripole part.

\subsection{Mode counting}

For completeness, let us also confirm in this gauge that the physical
modes are only two transverse modes with helicity $j=\pm 2$ and all the
other modes properly fall into the unphysical BRST quartets.

To do so, consider the free field equations of motion following from the
quadratic terms of $S_{\rm GR}=\int d^4x {\cal L}_{\rm GR}$,
which are also the equations of motion of the asymptotic fields:
\begin{eqnarray}
\frac{\delta S_{\rm GR}}{\delta\tlh^{\mu\nu}}:
&&\frac12 \square\tlh_{\mu\nu} -\partial_\rho\partial_{(\nu}\tlh_{\mu)}{}^\rho
-\frac14 \eta_{\mu\nu}\square\tlh
-\partial_{(\mu} b_{\nu)}+\frac12 \eta_{\mu\nu}b=0,
\label{eq:h} \\
\frac{\delta S_{\rm GR}}{\delta S}:&&\partial^\mu b_\mu=0, \label{eq:S}\\
\frac{\delta S_{\rm GR}}{\delta b_\mu}:&&\partial_\nu\tlh^{\mu\nu}-\partial^\mu S=0,
\label{eq:bmu}\\
\frac{\delta S_{\rm GR}}{\delta b}:&&\tlh=0,  \label{eq:b}\\
\frac{\delta S_{\rm GR}}{\delta \bar c_\mu}, \frac{\delta S_{\rm GR}}{\delta \bar d}:
&&\square c^\mu+\partial^\mu d=0,  \qquad \partial_\mu c^\mu=0,
\quad \rightarrow \quad \square d=0,
\label{eq:cd}\\
\frac{\delta S_{\rm GR}}{\delta c^\mu},
\frac{\delta S_{\rm GR}}{\delta d}:
&& \square\bar c_\mu+\partial_\mu\bar d=0,  \qquad \partial^\mu\bar c_\mu=0,
\quad  \rightarrow \quad  \square\bar d=0,
\label{eq:bar_cd}
\end{eqnarray}
where the bracket (\ ) attached to the indices means the weight 1
symmetrization; e.g., $A_{(\mu} B_{\nu)}
= (1/2)\big(A_{\mu} B_{\nu}+A_{\nu} B_{\mu}\big)$.
Taking the $\eta^{\mu\nu}$-trace of \p{eq:h} and using \p{eq:bmu} and \p{eq:b}, we find
\bea
-\square S + 2b =0\,.
\label{eq:Sb}
\eea
The divergence of the gravity equation (\ref{eq:h}), combined with \p{eq:S} -- \p{eq:b}, yields
\begin{eqnarray}
-\frac12( \square\partial_\mu S + \square b_\mu-\partial_\mu b)=0
\quad \rightarrow\quad  \square b_\mu+\partial_\mu b =0
\quad \rightarrow\quad  \square b =0\,,
\label{eq:Sbb}
\end{eqnarray}
where in the first step, we have used \p{eq:Sb}, and in the next step,
we took the divergence and used \p{eq:S}.
These equations (\ref{eq:Sb}) and (\ref{eq:Sbb}) imply that
$S$ and $b_\mu$ fields satisfy dipole equations
$\square^2 S=0$ and $\square^2 b_\mu=0$, and their dipole parts are supplied by
the simple pole $b$ field.
The gravity field equation (\ref{eq:h}) is rewritten by using
Eqs.~(\ref{eq:bmu}) and (\ref{eq:b}) into
\begin{equation}
\frac12\square\tlh_{\mu\nu} - \partial_\mu\partial_\nu S
-\partial_{(\mu} b_{\nu)}+\frac12 \eta_{\mu\nu}b=0.
\end{equation}
Applying $\square$ to this and using Eqs.~(\ref{eq:Sb}) and (\ref{eq:Sbb}),
we find
\begin{equation}
\frac12\square^2\tlh_{\mu\nu} - \partial_\mu\partial_\nu b =0.
\end{equation}
So, we see that $\tlh_{\mu\nu}$ field is now a tripole field, and the tripole part
is supplied by the simple pole $b$ field and the dipole parts are
supplied by simple pole parts of $S$ and $b_\mu$.

The conventional method to count the number of independent modes for such
a system containing multipole fields,
is to redefine such multipole fields into simple pole fields by subtracting
the mixed simple pole fields.
Since it is quite cumbersome to do so, however,
we here adopt the 4-dimensional Fourier expansion \cite{Nakanishi:1966zz} defined as
\begin{equation}
\phi(x) = \frac1{\sqrt{(2\pi)^3}} \int d^4p \,\theta(p^0) \left[
\phi(p)e^{ipx} + \phi^\dagger(p)e^{-ipx}\right].
\label{eq:2.35}
\end{equation}
Namely, we treat the multipole fields as they stand; the 4-dimensional
operators $\phi(p)$ and $\phi(p)^\dagger$ annihilate and create the
multipole particles as they stand.
The BRST singlet physical modes must of course be simple pole fields.
Multipole fields are necessarily unphysical and so will fall into
the BRST quartet. We will see
that the members of a BRST quartet have a common multipole structure
to decouple among them.
\def\mbf#1{{\boldsymbol #1}}
We note that, when $\phi(x)$ is a simple pole field,
$\phi(p)$ is given in terms of the usual
annihilation operator $\phi(\mbf p)$ by 3-dimensional Fourier transform as
\begin{equation}
\phi(p) = \theta(p^0) \delta(p^2) \sqrt{2|\mbf p|} \phi(\mbf p).
\end{equation}

Now the counting the number of modes by 4-dimensional Fourier modes
$\phi(p)$ and $\phi^\dagger(p)$ is very easy and the same as the number
of field's components. We have $10(\tlh_{\mu\nu})+ 1(S)+ 4(b_\mu)+ 1(b)=16$
component fields in the boson sector, and
$4(c^\mu)+1(d)=5$ and $4(\bar c_\mu)+1(\bar d)=5$ components in the ghost and antighost sectors, respectively.
But they are subject to the constraint equations of motion which reduce the number of independent modes.

First $\tlh_{\mu\nu}$ field is subject to 4 constraints by Eq.~(\ref{eq:bmu})
and 1 traceless constraint (\ref{eq:b}). Thus, $\tlh_{\mu\nu}$
contains $10 - 4 -1 =5$ independent modes.
The $b_\mu$ field obeys the transverse equation (\ref{eq:S}), so leaving
$4-1=3$ independent modes. $S$ and $b$ each gives an independent mode.
Similarly, in the ghost sector, $c^\mu$ and $\bar c_\mu$ fields each
have 3 independent modes by Eqs.~(\ref{eq:cd}) and (\ref{eq:bar_cd}) and $d$ and $\bar d$
each gives an independent mode.

Now recall the BRST transformation at the linearized level (which is also the BRST
transformation of the asymptotic fields under the perturbative assumption).
\begin{eqnarray}
[iQ_\B, \tlh_{\mu\nu}(p)] &=& -ip_\mu c_\nu(p)-ip_\nu c_\mu(p)
\quad (\because \partial_\rho c^\rho=0),
\label{eq:QBh}\\
{}\{iQ_\B, \bar c_\mu(p) \} &=& ib_\mu(p), \\
{}[iQ_\B, S(p)] &=& d(p), \qquad  \{iQ_\B, \bar d(p) \} = ib(p)\,.
\label{eq:BRStrf}
\end{eqnarray}

As for the 5 independent modes among 10 gravity fields $\tlh_{\mu\nu}$,
we first define the following 6 linear combinations of $\tlh_{\mu\nu}$,
choosing the $x^3$ axis along $\mbf p$:
\begin{eqnarray}
\tlh_{{\rm T}1}(p) &=& \frac12( \tlh_{11}(p)-\tlh_{22}(p) ), \nn
\tlh_{{\rm T}2}(p) &=& \tlh_{12}(p), \nn
\chi_0(p) &=& \frac1{2p_0}\left(\tlh_{00}(p)
-\frac12(\tlh_{11}(p)+\tlh_{22}(p))\right), \nn
\chi_i(p) &=& \frac1{p_0}\tlh_{i0}(p) \ \ \hbox{for} \ \ i=1,2, \nn
\chi_3(p) &=& \frac1{2p_3}\left(\tlh_{33}(p)
+\frac12(\tlh_{11}(p)+\tlh_{22}(p))\right).
\end{eqnarray}
These fields have very simple properties under the BRST transformation
(\ref{eq:QBh}): The transverse states are BRST singlets,
\begin{equation}
[iQ_\B, \tlh_{{\rm T}i}(p)] = 0 , \qquad (i=1,2),
\end{equation}
and $\chi_\mu(p)$'s satisfy
\begin{eqnarray}
[iQ_\B, \chi_\mu(p)] = -i c_\mu(p), \qquad (\mu=0,1,2,3).
\end{eqnarray}
They also satisfy the transversality condition $p^\mu\chi_\mu(p)=0$ so that they have only 3 components.
Together with $\tlh_{{\rm T}i}$, they give the 5 independent components from $\tlh_{\mu\nu}$.

Other than these 5 modes $\big(\tlh_{{\rm T}i}(p), \chi_\mu(p)\big)$ from $\tlh_{\mu\nu}$,
we only have 5 bosons $S(p), b_\mu(p), b(p)$ and 4 ghosts
$(c^\mu, d)$ plus 4 antighosts $(\bar c_\mu, \bar d)$.
We now see that aside from two physical transverse modes
$\tlh_{{\rm T}i}(p)$, all the other modes fall into
{\it eight} BRST doublets:
\begin{eqnarray}
\begin{array}{rcccl}
\hbox{vector BRST doublets} &:& \bigl( \chi^\mu, c^\mu \bigr)&;&
\bigl( \bar c_\mu, b_\mu \bigr)\,,\\
\hbox{scalar BRST doublets} &:& \bigl(  S,  d \bigr)&;& \bigl( \bar d, b \bigr)\,.
\end{array}
\end{eqnarray}
Recall that these vectors $\chi^\mu, c^\mu, \bar c_\mu, b_\mu$ are all transversal
and we count each of them as $d-1=3$ modes, so we have $(d-1)\times2+ 1\times2=8$ BRST doublets in total.
They indeed have the following BRST transformation laws:
\begin{equation}
\left[iQ_\B,
\begin{pmatrix}\chi^\mu \\ S \end{pmatrix} \right] =
\begin{pmatrix}-ic^\mu \\ d \end{pmatrix},
\qquad
\left[iQ_\B,
\begin{pmatrix}\bar c_\mu \\ \bar d \end{pmatrix} \right] = i
\begin{pmatrix} b_\mu \\  b \end{pmatrix} .
\end{equation}
These eight BRS doublets fall into {\it four} BRST quartets which decouple
from the physical subspace,
and we are left with the two physical transverse modes with helicity $j=\pm2$.

It is, however, worth noting that the metric structure of these
eight BRST doublets is slightly unfamiliar one. In particular, it is
{\it not correct} to say that the following pair of the BRST doublets
gives a {\it BRST quartet}:
\begin{equation}
\bigl[ (S(p), d(p));  (\bar d(p),  b(p)) \bigr].
\end{equation}
This is because $\VEV{ d(p) \bar d^\dagger(q) }$ ${}= -i \VEV{ S(p) b^\dagger(q) }=0$, as is seen shortly in
Eq.~(\ref{eq:dbard}), which implies that the BRST doublet $(S(p), d(p))$ does not have
nonvanishing inner-product with the other BRST doublet
$(\bar d(p), b(p))$ and so they do not constitute a BRST quartet
in the proper sense of the terminology.

The 4-dimensional commutation relations
for these asymptotic fields can be derived from the
expressions for the propagator
$i\times {\Gamma^{(2)}}^{-1}$ in
Eqs.~(\ref{eq:GammaInvBoson}) and (\ref{eq:GammaInvGhost})
by the standard procedure and are explicitly given in Eq.~(\ref{eq:4D-CR}) in the Appendix.
From Eq.~(\ref{eq:4D-CR}), we can see
the following commutation relations for the present independent modes:
\begin{eqnarray}
[\tlh_{{\rm T}i}(p), \tlh^\dagger_{{\rm T}j}(q)] &=& \delta_{ij} \theta(p^0) \delta(p^2) \delta^4(p-q),
\label{eq:tlhC}\\
{}[\tlh_{{\rm T}i}(p), \phi^\dagger(q)] &=& 0,  \quad \hbox{for}\ \
\phi=\chi_\mu, S, b_\mu, b,
\label{eq:tlhphi} \\
{} [\chi_\mu(p), b^\dagger_\nu(q)
] &=&
 \{ c_\mu(p), \bar c^\dagger_\nu(q) \}  \nn
&=& i \eta_{\mu\nu}\theta(p^0) \delta(p^2) \delta^4(p-q)
- i p_\mu p_\nu \theta(p^0) (-\delta'(p^2)) \delta^4(p-q),
\label{eq:cbarc} \\
{}[\chi_\mu(p), b^\dagger(q)] &=&
 \{ c_\mu(p), \bar d^\dagger(q)\}
= p_\mu \theta(p^0) \delta(p^2) \delta^4(p-q),
\label{eq:cbard}\\
{}[S(p), b^\dagger(q)] &=&
i \{ d(p), \bar d^\dagger(q) \}=0,
 \label{eq:dbard}
 \\
{}[S(p), b_\mu^\dagger(q)] &=&
i \{ d(p), \bar c_\mu^\dagger(q) \}
= -ip_\mu \theta(p^0) \delta(p^2) \delta^4(p-q).
\label{eq:dbarc}
\end{eqnarray}
The first Eq.~(\ref{eq:tlhC}) means that
the transverse modes $\tlh_{{\rm T}i}$ really are simple pole fields
and have positive norms, and
the second (\ref{eq:tlhphi}) show that
they are orthogonal to all the other modes.
Equation~(\ref{eq:dbard}) shows the vanishing inner-product
$\VEV{d(p) \bar d^\dagger(p)} =0$ mentioned above.
Equations~(\ref{eq:cbard}) and (\ref{eq:dbarc}) show the cross inner-product relations
between the scalar and vector BRST doublets; Eq.~(\ref{eq:cbard}) shows
that the partner BRST doublet of the scalar BRST doublet $(\bar d, b)$ is
the longitudinal component $(\chi_{\rm L}, c_{\rm L})$ of the vector
BRST doublet $(\chi_\mu, c_\mu)$ (Here, the longitudinal means the mode whose
polarization vector is proportional to $p^\mu$):
\begin{equation}
\VEV{ c_{\rm L}(p) \bar d^\dagger(q)}=
\VEV{ \chi_{\rm L}(p)\, Q_\B\, \bar d^\dagger(q)}=
\VEV{ \chi_{\rm L}(p) b^\dagger(q)} \neq 0 .
\label{eq:WTrelation}
\end{equation}
In the same way, Eq.~(\ref{eq:dbarc}) means that the BRST doublet
$(S(p), d(p))$ has nonvanishing innerproduct with the
$(\bar c_{\rm L}(p), b_{\rm L}(p))$.
Thus we can identify the following four BRST quartets (pairs of BRST doublets)
in the present case, in the proper sense of terminology of {\it BRST quartet}:
\begin{eqnarray}
\hbox{transverse quartets}\ (i=1,2)&:& (\chi_i(p), c_i(p)) \leftrightarrow (\bar c_i(p), b_i(p)), \nn
\hbox{longitudinal $\chi$ quartet}&:& (\chi_{\rm L}(p), c_{\rm L}(p)) \leftrightarrow (\bar d(p), b(p)), \nn
\hbox{longitudinal $\bar c$ quartet}&:& (S(p), d(p)) \leftrightarrow (\bar c_{\rm L}(p), b_{\rm L}(p)) ,
\end{eqnarray}
with $\leftrightarrow$ indicating the mutual partner doublets with which
nonvanishing inner-products exist like Eq.~(\ref{eq:WTrelation}).
These four sets of fields are the real BRST quartets and completely decouple
from the physical sector,
leaving only 2 transverse modes in GR.

\subsection{Cosmological constant in GR in the unimodular gauge}
\label{GRandUG}

Before closing this section, we should comment on the crucial
difference on the cosmological constant between GR in the unimodular gauge and UG.

Our total action~$\p{GRaction}+\p{GR:gf}$ in the present unimodular gauge GR
yields the field equation for the metric
\bea
&&R_{\mu\nu} -\frac12 g_{\mu\nu}(R-2\Lambda + b)
-(\partial_{(\mu}b_{\nu)}+b_{(\mu}\partial_{\nu)}S) +\frac12 g_{\mu\nu}
g^{\rho\sigma}(\partial_\rho b_\sigma+b_\rho\partial_\sigma S) \nn
&& \hspace{2em}{}+\mbox{(terms involving the FP ghost fields $c^\mu, \bar c_\mu, d, \bar d$)} = 0.
\label{eq:GRfe}
\eea
Let us consider the classical vacuum solution of this gravity equation
as the background field configuration of the quantum theory.
If we look for the solution with vanishing vector
NL field $b_\mu=0$, then this Eq.~\p{GRfe} reduces to the form
\bea
R_{\mu\nu} -\frac12 g_{\mu\nu}(R-2\Lambda + b) =0\,,
\label{GRfe}
\eea
which is identical to the field equation~\p{fe} in UG if we
identify $b$ here with the multiplier field $\lambda$ there.
One might then be tempted to conclude that the cosmological constant
in the action has no physical meaning in GR in the unimodular gauge, just like UG.

However, in GR, there is a crucial physical state condition \cite{KO1977}
\begin{equation}
Q_\B |{\rm phys}\rangle =0 ,
\end{equation}
and the NL field $b$ is a BRST exact field,  BRST transform of the
antighost field $\bar d$; $\{ Q_\B, \bar d \} = b$.
Since the vacuum must be a physical state, the NL field $b$ must always
have vanishing vacuum expectation value:
\begin{equation}
\VEV{ b(x) } = \VEV{ \{ Q_\B, \bar d(x) \} } = 0.
\end{equation}
This is also true for $b_\mu$.
At the tree level, the vacuum expectation value (VEV) of the field operator product equals
the product of the VEV of each field. So all the terms containing FP ghost fields vanish.
Therefore, the above field equation \p{eq:GRfe} correctly reproduces the classical Einstein equation
with the cosmological constant~\p{ein}.
This is to be expected since physical contents should be independent of the gauge fixing,
and in GR in covariant gauge like de Donder gauge, the cosmological constant
is really a physical quantity which determines the scalar curvature of the vacuum.

Then, what is the difference between UG and GR in the unimodular gauge?
It is the field properties of the multiplier field $\lambda$ in UG and the NL field $b$ in GR
that make the difference.
For the multiplier field $\lambda$ in UG, unlike $b$, there is no constraint that its VEV should vanish or any other
matrix element. So, we have the gravity field equation \p{fe} and $\lambda$ is a field that should be eliminated by
using the field equation. This requires the manipulation described in the introduction, and we can get
only the traceless part of the field equation~\p{ugfe}.
We see that $\la$ gets VEV to cancel the cosmological constant.

\section{Implications for Unimodular Gravity}

We would like to discuss if and how the above formulation of GR in the unimodular gauge may cast any light
on the problem of covariant quantization of UG with the action~\p{UGaction}.
One notable fact is that, as we have discussed in the preceding section,
the condition~\p{uni} is imposed as a gauge fixing in GR by one of the member $b$ of the BRST quartet
but here it is imposed as a constraint by the Lagrange multiplier field $\lambda$
independent of the BRST transformation.
This restricts the invariance of the theory to TDiff.

Since UG has fixed spacetime volume, it appears natural to consider that the usual GR is recovered by summing
over the spacetime volume in UG.
This reminds us of the suggestion made long time ago by Hawking that
the partition function for the theory with $\Lambda$ are related by ``Laplace transform''
with that with spacetime volume $V$ and vice versa~\cite{Hawking}.
In the Minkowski space, the transformation should be Fourier transformation.
So let us consider what happens if we make Fourier transformation of the partition function
of GR in the unimodular gauge.

The partition function for GR in the unimodular gauge is given by
\begin{eqnarray}
Z_{\rm GR}[\Lambda,\omega] &=& \!\!\int [D g_{\mu\nu}] [D S] [D b_\mu] [D c^\mu] [D\bar c_\mu]
[D d] [D \bar d] [Db]
\nonumber\\
&&\exp \left[i\frac1{16\pi G}\int d^4x\,
\left\{\sqrt{-g} (R-2\Lambda)+ {\cal L}_{\GF+\FP}\right\} \right].
\label{eq:ZGR}
\end{eqnarray}
If we multiply this by
\bea
\exp\left[\frac{i\Lambda}{8\pi G} \int d^4x \omega\right],
\eea
and integrate over $\Lambda$ first, we get $\d\left(\int d^4 x \sqrt{-g}-\int d^4 x\, \omega\right)$.
This is the global relation~\p{vol}, which is a consequence in UG,
but this does not immediately lead to the local constraint $\sqrt{-g}-\omega=0$.
This is close to the UG, and it may be instructive to study the theory with this condition
changed into a local one:
\bea
&& Z_{\rm UG} \equiv \int [D\la] \!\!\int [D g_{\mu\nu}] [D S] [D b_\mu] [D c^\mu] [D\bar c_\mu]
[D d] [D \bar d] [Db] \nn
&&\hs{20} \exp \left[i\frac1{16\pi G}\int d^4x\,
\left\{\sqrt{-g} R + {\cal L}_{\GF+\FP}+\la(\sqrt{-g}- \omega) \right\} \right].
\eea
The fact that we obtain this by summing over cosmological constant would be consistent with the fact
that the cosmological constant is not determined in UG.

Recall that we have the term of the same form $b(\sqrt{-g}-\omega)$ as this constraint in the gauge fixing
terms~\p{GR:gf},
and then this term may be absorbed into $\la$.
The integral over $b$ may be performed and yields a trivial constant
\begin{equation}
\int Db \exp \left[ \frac{i}{16\pi G} \int d^4x \frac{\beta}2 b^2 \right] =
{\rm const.}\prod_x \frac{1}{\sqrt{\beta}}.
\end{equation}
Absorbing this constant into the definition of the partition function, we find the transformed partition function
\bea
&&  Z_{\rm UG} = \!\!\int [D g_{\mu\nu}] [D S] [D b_\mu] [D c^\mu] [D\bar c_\mu] [D\la]
[D d] [D \bar d] \nn
&&\hs{20} \exp \left[i\frac1{16\pi G}\int d^4x\,
\left\{\sqrt{-g} R + {\cal L}_{\GF+\FP}'+\la(\sqrt{-g}-\omega) \right\} \right],
\label{quantumUG}
\eea
where
\bea
{\cal L}_{\GF+\FP}' &=& b_\mu \Bigl( \partial_\nu\tilde g^{\mu\nu} - \tilde g^{\mu\nu}\partial_\nu S \Bigr)
+ \frac{\a}{2}b_\mu^2 \nn
&&+i\bar c_\mu \Bigl( \partial_\nu\delta_\B(\tilde g^{\mu\nu}) - \delta_\B(\tilde g^{\mu\nu})\partial_\nu S
 - \tilde g^{\mu\nu}\partial_\nu d \Bigr) + i \bar d \left( \sqrt{-g}\nabla_\mu c^\mu \right).
\eea
This looks like the same as GR in the unimodular gauge.
Since it has the same structure as GR in the unimodular gauge,
one would naively expect that the analysis in the preceding section goes through and
the remaining degrees of freedom are the same as GR.
In fact, the Feynman rules for both theories are the same, and any scattering amplitudes calculated
in both theories agree.

Considering all these circumstances, it appears that this theory may be a possible candidate for
the quantum theory of UG.
Then what is the difference between GR in the unimodular gauge and this theory?
The difference lies in the fundamental difference in the nature of the field $b$ and $\lambda$,
as we have discussed in sect.~\ref{GRandUG}.
In GR, the field $b$ giving the unimodular constraint is the NL field, a member of BRST quartet,
and the theory has the BRST invariance. As a consequence, it does not have VEV.
In contrast, in the current theory, $\la$ is BRST singlet and can have VEV,
and may cancel the cosmological constant.
However, since the action in \p{quantumUG} is no longer invariant under the BRST transformation
because $\lambda\neq b$, we do not have the same physical state condition as in GR.
If we defined another new BRST transformation under which our $\lambda$ field
were the BRST transform of the antighost $\bar d$,
$\{ Q_\B, \bar d\} = \lambda$, and then we would have the physical state condition but
the VEV of $\la$ would have to vanish, reproducing GR in the unimodular gauge.

As stated above, we have the same Feynman rules and get the same amplitudes in both theories.
For the transverse gravitons, these give amplitudes without any problem.
The problem manifests itself when we calculate the amplitudes for the longitudinal modes.
In the BRST invariant theory, the physical state condition forbids them to appear by themselves
but must come together with the FP ghosts, and their contributions vanish.
Without the physical state condition, nothing tells us such amplitudes should be absent.

Another point to be noticed is that in the present theory, the invariance under TDiff
\bea
\d_B g_{\mu\nu}=\nabla_\mu c_\nu^\T + \nabla_\nu c_\mu^\T, \qquad
\nabla^\mu c_\mu^\T=0,
\label{onshells}
\eea
is realized only on shell. We can see that this constraint is imposed by the field equations for
the ghosts $d$ and $\bar d$.
If one started with the UG with the action~\p{UGaction2},
we would have the invariance only under \p{onshells} off shell, but then we could not see the necessity
of introduction of the additional fields $(S, d, \bar d)$, which are necessary for the decoupling
of the modes other than those in GR.
It is plausible that the formulation of quantum UG must be realized by fixing only TDiff.
This should be related to the question how to specify the physical states in this theory.

To summarize, as far as the Feynman rules and other properties are concerned, both theories describe the same
theory except for the treatment of the cosmological constant,
but here the criterion for choosing our {\it physical asymptotic states} is lacking in the above theory.
We may get the same scattering amplitudes among the same set of particles in both theories,
but it is not specified in this theory which scattering amplitudes are physical.
These problems of identifying physical asymptotic states and gauge fixing the invariance under TDiff off shell
in the theory are under study. We hope to report on this in the near future.
\\

\section*{Acknowledgment}

We thank Roberto Percacci for valuable discussions.
T.K. is supported in part by the JSPS KAKENHI Grant No. JP18K03659.
N.O. is supported in part by the Grant-in-Aid for Scientific Research Fund of the JSPS (C) No. 16K05331,
No. 20K03980, and Taiwan MOST No. 110-2811-M-008-510.

\appendix

\section{4-dimensional commutation relations from propagators}

The propagators, or the inverse of the 2-point vertices
$i\times {\Gamma^{(2)}}^{-1}$ of the boson fields and FP ghost fields
in the present system are computed in
Eqs.~(\ref{eq:GammaInvBoson}) and (\ref{eq:GammaInvGhost}), respectively.
We can rewrite those expression
for
$\Propag{ \phi_i(x) \phi_j(0) } = i\times {\Gamma^{(2)}}_{ij}^{-1}$
in $x$-space in terms of the invariant functions
\begin{eqnarray}
D_{F}(x) &=& \Delta_{F}(x;m^2)\bigr|_{m^2=0} \nn
E_{F}(x) &=& -\frac{\partial}{\partial m^2}\Delta_{F}(x;m^2)\bigr|_{m^2=0} \nn
F_{F}(x) &=& \frac12\left(\frac{\partial}{\partial m^2}\right)^2
\Delta_{F}(x;m^2)\bigr|_{m^2=0} \\
\Delta_F(x;m^2) &=& \int \frac{d^4p}{i(2\pi)^4} \frac{e^{ipx}}{m^2+p^2-i\e} \nn
\Delta(x;m^2) &=& \int \frac{d^4p}{i(2\pi)^3} \epsilon(p^0) \delta(m^2+p^2)e^{ipx}
\end{eqnarray}
with suffix $F$ meaning Feynman's causal functions.
Then, for any free theory, the commutation relations $[\phi_i(x), \phi_j(0)]$
can be obtained simply by replacing the invariant functions by the same
invariant functions without the suffix $F$ multiplied by $i$; that is,
replacing $\Delta_F(x;m^2)$ by $i\Delta(x;m^2)$, $D_F(x)$ by $iD(x)$ etc.
(This rule actually holds even for interacting Heisenberg fields if
we use the spectral function representation.)

Applying this rule to the $x$-space representation of the
propagators obtained in
Eqs.~(\ref{eq:GammaInvBoson}) and (\ref{eq:GammaInvGhost}),
we find the following 4-dimensional commutation relations for our
asymptotic fields:
\begin{eqnarray}
[\tlh_{\mu\nu}(x), \tlh_{\rho\sigma}(y)]
&=& (\eta_{\mu\rho}\eta_{\nu\sigma}+
\eta_{\mu\sigma}\eta_{\nu\rho} -\eta_{\mu\nu}\eta_{\rho\sigma}) iD(x-y)  \nn
&&{}+\{ {\cal A}_{\mu\nu,\rho\sigma}-2{\cal B}_{\mu\nu,\rho\sigma} \}iE(x-y)
-4\partial_\mu\partial_\nu\partial_\rho\partial_\sigma iF(x-y) \nn
{}[\tlh_{\mu\nu}(x), S(y)]
&=&
\eta_{\mu\nu}iD(x-y) +4 \partial_\mu\partial_\nu iE(x-y), \nn
{}[\tlh_{\mu\nu}(x), b_\rho(y)]
&=&
\left(\eta_{\mu\rho}\partial_\nu+ \eta_{\nu\rho}\partial_\mu\right) iD(x-y)
+2 \partial_\mu\partial_\nu\partial_\rho iE(x-y), \nn
{}[\tlh_{\mu\nu}(x), b(y)]
&=&
-2\partial_\mu\partial_\nu iD(x-y), \nn
{}[S(x), S(y)]&=& -3iD(x-y), \nn
{}[S(x), b_\rho(y)]&=& -\partial_\rho iD(x-y), \nn
{}[b_\mu(x), b_\nu(y)]&=& [b_\mu(x), b(y)] = [b(x), b(y)] =0 \nn
\{c_\mu(x), \bar c_\nu(y)\} &=&
-\eta_{\mu\nu} D(x-y) - \partial_\mu\partial_\nu E(x-y), \nn
\{c_\mu(x), \bar d(y)\} &=&
\partial_\mu D(x-y), \nn
\{ d(x), \bar c_\mu(y) \} &=&
-\partial_\mu D(x-y), \nn
\{ d(x), \bar d(y) \} &=& 0,
\end{eqnarray}
where
\begin{eqnarray}
{\cal A}_{\mu\nu,\rho\sigma}&=&
\eta_{\mu\rho}\partial_\nu\partial_\sigma+\eta_{\mu\sigma}\partial_\nu\partial_\rho
+\eta_{\nu\rho}\partial_\mu\partial_\sigma+\eta_{\nu\sigma}\partial_\mu\partial_\rho ,
\nn
{\cal B}_{\mu\nu,\rho\sigma}
&=&
\eta_{\mu\nu}\partial_\rho\partial_\sigma+\eta_{\rho\sigma}\partial_\mu\partial_\nu .
\end{eqnarray}
The Fourier transforms of these give the commutation relations
of the `creation-annihilation' operators $\phi(p)$ and $\phi^\dagger(p)$ defined by
4-dimensional Fourier expansion in Eq.~(\ref{eq:2.35}). We find
\begin{eqnarray}
[\tlh_{\mu\nu}(p), \tlh^\dagger_{\rho\sigma}(q)]
&=& (\eta_{\mu\rho}\eta_{\nu\sigma}+
\eta_{\mu\sigma}\eta_{\nu\rho} -\eta_{\mu\nu}\eta_{\rho\sigma}) \theta(p^0) \delta(p^2) \delta^4(p-q)  \nn
&&{}+\{ {\cal A}_{\mu\nu,\rho\sigma}\bigr|_p
-2{\cal B}_{\mu\nu,\rho\sigma}\bigr|_p \}
\theta(p^0) (-\delta'(p^2)) \delta^4(p-q) \nn
&&{}
-2 p_\mu p_\nu p_\rho p_\sigma \theta(p^0) \delta''(p^2) \delta^4(p-q), \nn
{}[\tlh_{\mu\nu}(p), S^\dagger(q)]
&=&
\eta_{\mu\nu}\theta(p^0) \delta(p^2) \delta^4(p-q) -4 p_\mu p_\nu
\theta(p^0)(-\delta'(p^2)) \delta^4(p-q), \nn
{}[\tlh_{\mu\nu}(p), b^\dagger_\rho(q)]
&=&
\left(\eta_{\mu\rho}ip_\nu+ \eta_{\nu\rho}ip_\mu\right) \theta(p^0) \delta(p^2) \delta^4(p-q)
-2i p_\mu p_\nu p_\rho \theta(p^0) (-\delta'(p^2)) \delta^4(p-q), \nn
{}[\tlh_{\mu\nu}(p), b^\dagger(q)]
&=&
 2 p_\mu p_\nu \theta(p^0) \delta(p^2) \delta^4(p-q), \nn
{}[S(p), S^\dagger(q)]&=& -3\theta(p^0) \delta(p^2) \delta^4(p-q), \nn
{}[S(p), b^\dagger_\rho(q)]&=& -ip_\rho\theta(p^0) \delta(p^2) \delta^4(p-q), \nn
{}[b_\mu(p), b^\dagger_\nu(q)]&=& [b_\mu(p), b^\dagger(q)]
= [b(p), b^\dagger(q)] =0 \nn
\{c_\mu(p), \bar c^\dagger_\nu(q)\}
&=&
i\eta_{\mu\nu} \theta(p^0) \delta(p^2) \delta^4(p-q) -ip_\mu p_\nu \theta(p^0) (-\delta'(p^2)) \delta^4(p-q), \nn
\{c_\mu(p), \bar d^\dagger(q)\}
&=&
p_\mu \theta(p^0) \delta(p^2) \delta^4(p-q), \nn
\{ d(p), \bar c^\dagger_\mu(q) \}
&=&
-p_\mu \theta(p^0) \delta(p^2) \delta^4(p-q), \nn
\{ d(p), \bar d^\dagger(q) \} &=& 0,
\label{eq:4D-CR}
\end{eqnarray}
where $ X\bigr|_p$ indicates that all the derivative factors
$\partial_\mu$ contained in
$X$ should be replaced by $ip_\mu$.


\end{document}